\documentclass[conference]{IEEEtran}
\IEEEoverridecommandlockouts
\usepackage{cite}
\usepackage{amsmath,amssymb,amsfonts}
\usepackage{algorithmicx,algpseudocode}
\usepackage[framemethod=tikz]{mdframed}
\usepackage{listings}
\usepackage{graphicx}
\usepackage{textcomp}
\usepackage{xcolor}
\usepackage{hyperref}       % hyperlinks
\usepackage{url}            % simple URL typesetting
\def\BibTeX{{\rm B\kern-.05em{\sc i\kern-.025em b}\kern-.08em
    T\kern-.1667em\lower.7ex\hbox{E}\kern-.125emX}}

\lstdefinestyle{code_style}{
    basicstyle=\ttfamily\footnotesize,
    breakatwhitespace=false,         
    breaklines=true,                 
    captionpos=b,                    
    keepspaces=true,                
    numbersep=5pt,                  
    showspaces=false,                
    showstringspaces=false,
    showtabs=false
}
\surroundwithmdframed[
  hidealllines=true,
  innerleftmargin=15pt,
  innertopmargin=0pt,
  innerbottommargin=0pt]{lstlisting}

\newcommand{\SL}[1]{~\S\ref{sec:#1}}

\newcommand{\SLP}[1]{~(\S\ref{sec:#1})}

\newcommand*\circled[1]{~\tikz[baseline=(char.base)]{
  \node[shape=circle,draw,inner sep=2pt] (char) {#1};}}

\begin{document}

\title{Developing Distributed High-performance Computing Capabilities of an Open Science Platform for Robust Epidemic Analysis \\

{\footnotesize
%% \textsuperscript{*}Note: Sub-titles are not captured in Xplore and should not be used
\thanks{This material is based upon work supported by the National Science Foundation under Grant No. 2200234, the National Institutes of Health under grant R01DA055502, the U.S. Department of Energy, Office of Science, under contract number DE-AC02-06CH11357, and the DOE Office of Science through the Bio-preparedness Research Virtual Environment (BRaVE) initiative.}
}}

\author{\IEEEauthorblockN{Nicholson Collier}
\IEEEauthorblockA{\textit{Decision and Infrastructure Sciences} \\
\textit{Argonne National Laboratory}\\
Lemont, IL, U.S.A. \\
% \orcidlink{0000-0002-2376-4156} \url{https://orcid.org/0000-0002-2376-4156}}
\url{https://orcid.org/0000-0002-2376-4156}}
\and
\IEEEauthorblockN{Justin M. Wozniak}
\IEEEauthorblockA{\textit{Data Science and Learning} \\
\textit{Argonne National Laboratory}\\
Lemont, IL, U.S.A. \\
\url{https://orcid.org/0000-0002-2441-2048}}
\and
\IEEEauthorblockN{Abby Stevens}
\IEEEauthorblockA{\textit{Decision and Infrastructure Sciences} \\
\textit{Argonne National Laboratory}\\
Lemont, IL, U.S.A. \\
\url{https://orcid.org/0000-0003-1976-1806}}
\and
\IEEEauthorblockN{Yadu Babuji}
\IEEEauthorblockA{\textit{Computer Science} \\
\textit{University of Chicago}\\
Chicago, IL, U.S.A. \\
\url{https://orcid.org/0000-0002-9162-6003}}
\and
\IEEEauthorblockN{Micka\"{e}l Binois}
\IEEEauthorblockA{\textit{Acumes project-team} \\
\textit{Inria centre at Université Côte d'Azur}\\
Sophia Antipolis, France \\
\url{https://orcid.org/0000-0002-7225-1680}}
\and
\IEEEauthorblockN{Arindam Fadikar}
\IEEEauthorblockA{\textit{Decision and Infrastructure Sciences} \\
\textit{Argonne National Laboratory}\\
Lemont, IL, U.S.A. \\
\url{https://orcid.org/0000-0001-7396-0350}}
\and
\IEEEauthorblockN{Alexandra W\"urth}
\IEEEauthorblockA{\textit{Acumes project-team} \\
\textit{Inria centre at Université Côte d'Azur}\\
Sophia Antipolis, France \\
\url{https://orcid.org/0000-0002-6099-0531}}
\and
\IEEEauthorblockN{Kyle Chard}
\IEEEauthorblockA{\textit{Computer Science} \\
\textit{University of Chicago}\\
Chicago, IL, U.S.A. \\
\url{https://orcid.org/0000-0002-7370-4805}}
\and
\IEEEauthorblockN{Jonathan Ozik}
\IEEEauthorblockA{\textit{Decision and Infrastructure Sciences} \\
\textit{Argonne National Laboratory}\\
Lemont, IL, U.S.A. \\
\url{https://orcid.org/0000-0002-3495-6735}}
}

\maketitle

\pagestyle{plain}

\begin{abstract}
% Might as well fill the column.
COVID-19 had an unprecedented impact on scientific collaboration.  The pandemic and its broad response from the scientific community has forged new relationships among domain experts, mathematical modelers, and scientific computing specialists.  Computationally, however, it also revealed critical gaps in the ability of researchers to exploit advanced computing systems.  These challenging areas include gaining access to scalable computing systems, porting models and workflows to new systems, sharing data of varying sizes, and producing results that can be reproduced and validated by others.
Informed by our team's work in supporting public health decision makers during the COVID-19 pandemic and by the identified capability gaps in applying high-performance computing (HPC) to the modeling of complex social systems, we present the goals, requirements, and initial implementation of OSPREY, an open science platform for robust epidemic analysis. The prototype implementation demonstrates an integrated, algorithm-driven HPC workflow architecture, coordinating tasks across federated HPC resources, with robust, secure and automated access to each of the resources. We demonstrate scalable and fault-tolerant task execution, an asynchronous API to support fast time-to-solution algorithms, an inclusive, multi-language approach, and efficient wide-area data management. The example OSPREY code is made available on a public repository.    

\vspace{0.3in}
\end{abstract}

\begin{IEEEkeywords}
high-performance computing, computational epidemiology, HPC workflows, open science platform
\end{IEEEkeywords}

\section{Introduction}

%\begin{enumerate}
%    \item motivation
%    \item OSPREY
%    \item Requirements for OSPREY
%\end{enumerate}
The societal importance of simulating social systems has been brought into sharper focus as a result of the COVID-19 pandemic. The pandemic has shown how epidemiologic modeling can inform decision-making in times of crisis and uncertainty, while also highlighting areas that must be addressed to create sustainable, efficient, and more effective approaches to understanding the relevant and complex social processes for biopreparedness and response. An area of particular interest has been in using high-performance computing (HPC) to calibrate and apply epidemiologic models for producing outputs such as forecasts of cases, resource needs, and disease outcomes, to gain insight into the evolving nature of the pandemic~\cite{Hotton2022Impact} and to provide quick turnaround decision support for public health stakeholders~\cite{Ozik2021population}.
% HPC approaches are ad hoc, inefficient, manual, and error prone.
% However, despite the unprecedented production~\cite{Else2020How,Cai2021International} and co-production of scientific work~\cite{ray_ensemble_2020,borchering_modeling_2021}, individual research groups have generally worked independently, necessitating a  large amount of heroic, overlapping or duplicative work that occurs within individual research groups seeking to exploit~\cite{Ozik2021population} (or attempt to exploit) advances in HPC, data management, machine learning (ML), artificial intelligence (AI), and automation methods when developing, calibrating, modifying, verifying, and validating the epidemiologic models. Further, differences across HPC resources can result in one-off calibration workflows, developed to meet the requirements of a specific HPC environment.

However, despite the unprecedented production~\cite{Else2020How,Cai2021International} and co-production of scientific work~\cite{ray_ensemble_2020,borchering_modeling_2021}, in the form of large ensemble forecasts and scenario modeling, individual research groups have generally worked independently, as they've sought to exploit~\cite{Ozik2021population} (or attempt to exploit) advances in HPC, data management, machine learning (ML), artificial intelligence (AI), and automation methods when developing, calibrating, modifying, verifying, and validating the epidemiologic models. This has necessitated a large amount of heroic, overlapping work, with results that risk lacking robustness, security, scalability, or efficiency. Further, differences across HPC resources can result in one-off workflows, developed to meet the requirements of a specific HPC environment.

To add to the complexity, the data upon which modelers have relied is heterogeneous, changing, and incomplete, requiring complex integration across diverse and novel surveillance signals. These features present significant challenges for use in calibrating epidemiologic models. At the COVID-19 pandemic onset, data were generally limited to diagnosed cases, deaths, and sometimes hospitalizations favoring simpler models that did not require estimation of many parameters. As the pandemic evolved, multiple new high-resolution data streams emerged at the local level allowing the models to capture dynamic features of the epidemic and adapt towards better realism. Anticipating the evolution of data streams would help develop flexible models that could be quickly scaled up in complexity and ingest dynamically changing data that vary between locations; however, data management infrastructure is needed to efficiently support this.

Further, research groups have often focused on a single modeling scope, with one modeling method (compartmental, meta-population, agent-based), geographical extent (city, county, state, national), and temporal scale (short-term forecast, medium/long-term planning). Even when multiple scopes are considered, they are rarely integrated into multi-resolution ensembles that can mutually inform, and which could be combined to rapidly support decision making during different stages of an unfolding public health emergency.

To begin addressing these issues, we present the design and initial implementation of the Open Science Platform for Robust Epidemic analYsis, or OSPREY, which seeks to lower the barriers to and automate epidemiologic model analyses, monitoring, and rapid response on HPC resources. We begin by describing OSPREY's goals and requirements in section \ref{sec:goals-reqs}, and then discuss related work in section \ref{sec:related-work}. Section \ref{sec:osprey-architecture} presents our prototype OSPREY HPC architecture and section \ref{sec:apis} describes the programming model and application programming interfaces (APIs) for working with the architecture components. Section \ref{sec:results} provides results from an example optimization workflow built using the prototype architecture and APIs. We summarize our contributions and discuss
future work directions in section \ref{sec:conclusion}.

\section{OSPREY Goals and Requirements}
\label{sec:goals-reqs}
\subsection{OSPREY Goals}

Here we describe the three key OSPREY capabilities that drive its design.

\subsubsection{Integrated, algorithm-driven HPC workflows}
OSPREY needs to facilitate access to HPC resources for epidemiologic model developers and for developers of model exploration algorithms (e.g., calibration, data assimilation, uncertainty quantification, optimization algorithms). HPC workflow software will connect data, simulation, and algorithmic tasks, and provision heterogeneous resources (CPU, GPU, and other specialized accelerators) matched to task types, such as simulation-intensive (CPU) or machine learning model-intensive (GPU). Workflows driven by complex, iterating, and asynchronous algorithms must be executable at scales ranging from HPC clusters to the largest supercomputers~\cite{Ozik2021population,Babuji2020Targeting}, while enabling modeling groups with different computing allocations and resource requirements to transparently utilize and build upon the complete suite of OSPREY capabilities by plugging code into an automated, scalable, and reusable framework.

\subsubsection{Data ingestion, curation, and management}
OSPREY will need to enable continuously running data assimilation analyses for melding data streams with up-to-date model forecasts. The platform will ingest real-time and near real-time data streams; curate, store, and index data; and manage epidemiologic models and model outputs. Data pipelines will quantify and adjust for data limitations and track data provenance. 

\subsubsection{Shared Development Environment (SDE) for rapid response and collaboration}
The OSPREY design is a response to our experiences in utilizing HPC resources in support of public health decision-making during the COVID-19 pandemic.
Computational experimentation, verification, and validation are collaborative activities that require rapid cooperative development as public health crises evolve.  Quickly and correctly porting various modeling and analysis codes to HPC resources is also an important activity that requires a range of expertise.  Thus, the OSPREY design will make available shared, flexible, automated, and scalable approaches to support rapid model exploration in a Shared Development Environment (SDE).

\subsection{OSPREY Requirements}
Next we describe the requirements needed for OSPREY to address the goals outlined above. 
First we present the main focus for this paper: requirements for integrated, algorithm-driven HPC workflows. The final two sections (\ref{sec:req-data},\ref{sec:req-sde}) describe other requirements related to  ``Data ingestion, curation, and management'' and ``Shared Development Environment (SDE) for rapid response and collaboration'' that are not the focus of this work, but we include for completeness.

\subsubsection{Integrated, algorithm-driven HPC workflows}
\label{sec:integ-hpc}

\paragraph{Coordinated multi-resource task execution} The breadth of model, simulation, and analyses employed in OSPREY necessitates a flexible approach to computation in which a range of computational tasks can be executed across a distributed ecosystem of heterogeneous remote resources.  Such tasks are varied and may include single-core tasks to multi-node MPI tasks, running on CPUs to AI accelerators, and requiring different performance guarantees (e.g., response time) to coordinate execution of complex workflows.

\paragraph{Robust, secure, and automated access to federated HPC resources}
\label{sec:federated} 
HPC and supercomputing resources have a wide range of access, authentication, and security protocols. OSPREY will need to allow for robust, secure, and automated access to computational resources. 

\paragraph{Scalable, fault-tolerant task execution}
Computational demand in epidemiologic workflows can vary dramatically over time. Furthermore, computational availability can fluctuate due to demand, resource priorities, and site specific preemption protocols. Being able to dynamically scale up (or down) the resources available for task execution, while retaining the integrity of the overall workflows is critical. 

\paragraph{Fast time-to-solution workflows}
\label{sec:fast-time}
Epidemiologic models have become essential tools for understanding and characterizing disease evolution. However, for them to be useful as decision support tools, they need to provide actionable insights quickly. There is, therefore, a need to develop and apply algorithms focused on fast time-to-solution, whether for model calibration, optimization, or other exploration. These include efficient, surrogate-based multi-objective optimization algorithms that can exploit the concurrency of HPC resources~\cite{Ozik2021population,Binois2021portfolio}, and asynchronous algorithms that can incrementally learn and adjust as new information is obtained. Being able to leverage the latest advances in data analytic and statistical libraries is desirable. OSPREY needs to provide the ability to effectively integrate and exploit such algorithms. 

\paragraph{Multi-language workflows}
Computational epidemiology is inherently cross disciplinary, bringing together a wide array of expertise from many domains. There is, therefore, not a single \emph{lingua franca} that can be assumed for developing the model exploration algorithms that drive the workflows. OSPREY will need to be inclusive and provide multi-language APIs for workflows.

\paragraph{Efficient wide-area data staging}
Data is fundamental to epidemiologic analyses, whether in the form of traditional data structures or ML/AI/mechanistic model artifacts. Workflow computations require efficient staging of these data elements and need to provide uniform access to them across the heterogeneous computing infrastructure. 

% The next two sections (\ref{sec:req-data},\ref{sec:req-sde}) describe the requirements under the ``Data ingestion, curation, and management'' and ``Shared development environment (SDE) for rapid response and collaboration'' OSPREY capabilities and, while they are not the focus of this paper, we include them for completeness.

\subsubsection{Data ingestion, curation, and management}
\label{sec:req-data}
OSPREY will need to provide reusable services to make it possible to quickly develop and deploy workflows based on real-world data streams and model artifacts on HPC resources.

\paragraph{Data stream ingestion}
Incoming data streams relevant to OSPREY workflows vary widely in type and size. OSPREY will need to develop flexible techniques to move and track data sets from their origin of publication, such as a city or health department portals, to their site of use, such as a HPC cluster or supercomputer, to the models they are used in. 

\paragraph{Automated data curation}
To address the heterogeneous, changing, and incomplete data typical of surveillance and other public health data that OSPREY workflows will utilize, there is a need to develop capabilities for creating data analysis pipelines, such as for data de-biasing, data integration, uncertainty quantification, and more general metadata and provenance tracking.

\paragraph{Managing algorithm and model artifacts}
Algorithm and model artifacts, such as model exploration state or calibrated model checkpoints, can be complex, large, and numerous and not local to a specific resource. OSPREY needs to manage these artifacts, and their associated metadata. Capabilities should allow model exploration algorithms to be easily rerun or continued, either on the original set of computing resources or different ones. Model checkpoints should be easily selected, staged for execution, and run.

\subsubsection{SDE for rapid response and collaboration}
\label{sec:req-sde}
OSPREY will make available a Shared Development Environment (SDE) to make it possible to quickly share, validate, and scale models and workflows on HPC resources.  A notable feature of the OSPREY SDE is that it is not based on hardware or Infrastructure-As-A-Service (IaaS) products, but rather on portable workflows that run on the federated HPC systems~\SLP{federated} to which the user already has access.  

\paragraph{Model and workflow sharing}
Sharing scientific workflows is known to be difficult due to the typical intricacies of interactions with complex HPC systems components including shared filesystems, schedulers, and various environmental settings.  The standardized OSPREY workflow structure and its use of portable tools will make it more likely that ``works for me'' also means it will ``work for you,'' at least at the systems level.

\paragraph{Model validation and publishing}
At the scientific level, models must be calibrated, validated, and published using well-defined data sets.  We will employ best practices from the DevOps ecosystem~\cite{7140503} to make it easier for modelers to post complete models with the data used to validate them for reproduction, extension, or scaling by others, with the capability to detect correctness regressions.

\section{Related Work}
\label{sec:related-work}

The desire to perform secure remote computation, particularly for scientific computations, is not new, with methods such as SSH commonly used to securely
execute commands over insecure network connections. 
Recent paradigms, such as Grid computing and cloud computing, have introduced more seamless methods
for remote computing via common APIs, authentication frameworks, and infrastructure abstractions. 

The Globus Grid Resource Allocation Manager (GRAM)~\cite{Feller2007GT4GA} defined a common, web-accessible API for interacting with various batch schedulers. Computing facilities, such as NERSC and TACC, are developing REST APIs to, amongst other things, submit and manage batch jobs (e.g., SuperFacility~\cite{enders20superfacility} and Tapis~\cite{stubbs21tapis}). 

Science gateways~\cite{Gateways_WWW} provide web-based interfaces to 
various science tools, resources, and data. 
They have become popular as they abstract the complexity of dealing with remote environments, consolidate the tools and resources needed for a specific application, and remove the need to use several interfaces. 
Gateways are typically developed to support a specific
domain or application; however, many build 
upon general-purpose frameworks, such as Apache Airavata~\cite{marru11airavata}, Tapis, and Globus~\cite{Chard2014Efficient} that provide 
a range of data and compute management services.

Workflow management systems, such as Pegasus~\cite{deelman19pegasus}, Parsl~\cite{Babuji2019Parsl:}, Swift~\cite{zhao07swift}, and Balsam~\cite{salim19balsam}, are designed
to simplify the development and execution of 
sophisticated workflows---tasks defined and specified to be performed in a partial order with concurrent semantics. Workflow management systems enable
workflows to be executed on various computing resources
including clouds, grids, and HPC clusters. These systems typically are deployed locally on a compute resource, and use pilot jobs to manage execution of many tasks within a batch scheduler job. Some workflow management systems support multi-site execution; however, this typically relies on establishing (and maintaining) SSH connections or opening ports to shared databases or controllers. 

There is growing interest in remote computation 
as part of AI-based workflows. For example, Colmena~\cite{ward21colmena} is a Python-based framework designed to steer computational campaigns by enabling developers to wrap various fidelity tasks (e.g., simulations) and define functions to select which tasks to be executed next. Colmena supports the use of both Parsl and funcX~\cite{Chard2020funcX:} \SLP{tasks} to manage execution of tasks on HPC resources.

% \begin{enumerate}
%     \item Balsam
%     \item Colmena
%     \item (Possibly Pegasus/Condor)
%     \item (Big applications/science portals/gateways)
% \end{enumerate}

% \subsection{Systems used by OSPREY}

% \textbf{funcX} is a federated function-as-a-service (FaaS) platform that allows users to reliably, securely, and efficiently execute functions across computing resources. funcX adapts the cloud FaaS model, by retaining a single, highly-available cloud-hosted management service and combining it with an ecosystem of user-deployed funcX endpoints. Users register functions with the cloud, providing the function body; later they may then invoke that function by supplying input arguments and selecting an endpoint for execution. funcX endpoints can be deployed on various compute resources (e.g., clouds, clusters, Kubernetes) and are capable of elastically provisioning resources for function execution. Unlike workflow systems, the funcX endpoint interacts only with the funcX cloud service, using RabbitMQ to retrieve tasks and return results. Thus, users do not need to manually manage connections to remote computing resources. 

% \begin{enumerate}
%   \item{funcX}
%   \item{Swift/T}
% \end{enumerate}

\section{Prototype OSPREY HPC Architecture}
\label{sec:osprey-architecture}
%% What We Did (And How We Did It)

\begin{figure}[htbp]
    % \centerline{\includegraphics[width=0.5\textwidth]{figures/HPC_v6.png}}
  %% OSPREY-Arch-ME.pdf is at:
  %% https://docs.google.com/drawings/d/1BOfdyXkPckx85SulrOLY062lpTeDxA6Xv6AKu56q9Fc
  \centerline{\includegraphics[width=0.5\textwidth]{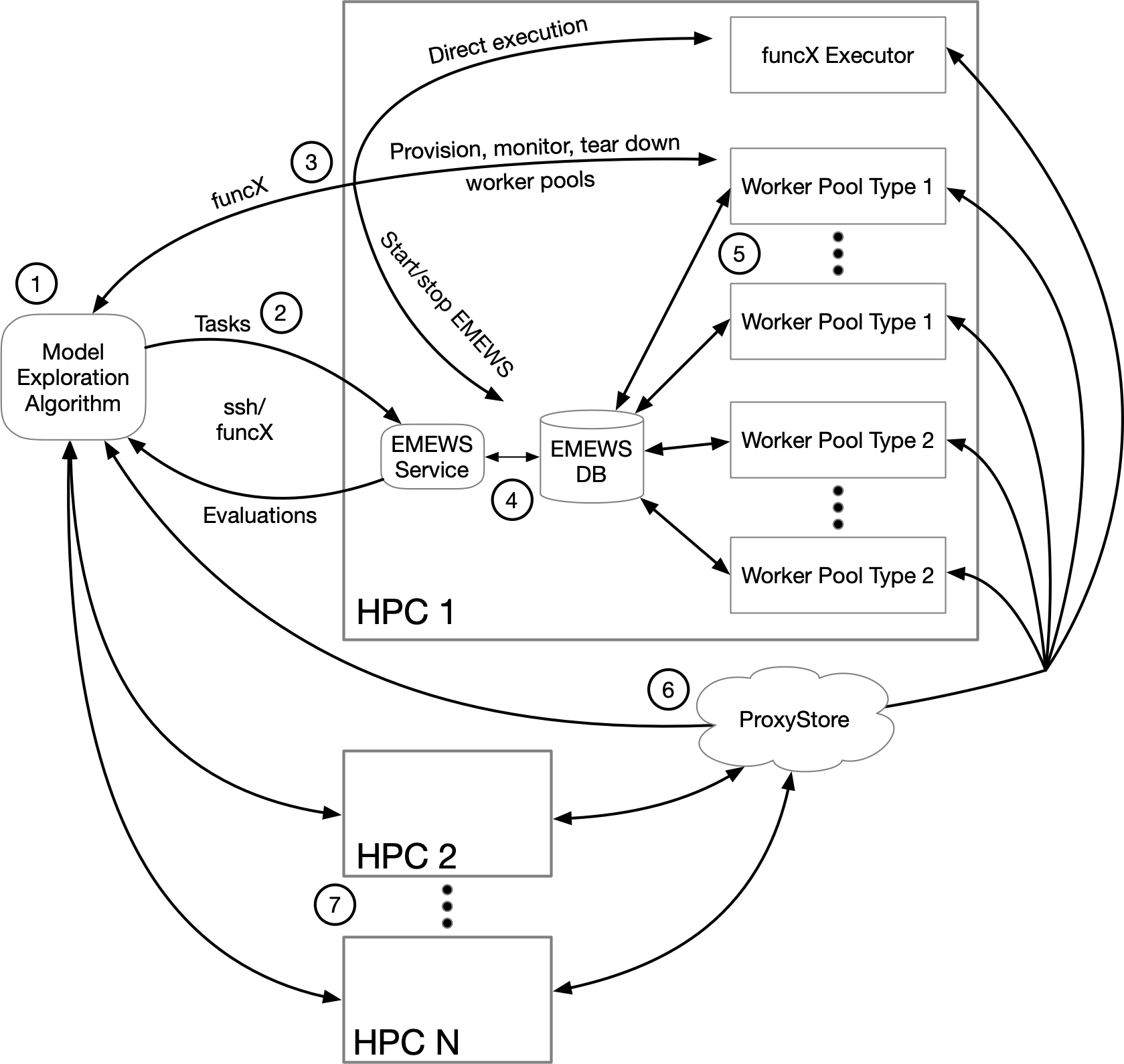}}
    \caption{Overview diagram of OSPREY prototype functionality.}
    \label{fig:overview}
    \end{figure}

In this section, we describe the prototype OSPREY HPC architecture, which we generalize from the EMEWS framework~\cite{Ozik2016From}. %that we built.  
The architecture is depicted in Figure~\ref{fig:overview}.  The architecture consists of the algorithm that controls the overall workflow\SLP{algmodule}, a control and task distribution system\SLP{tasks}, a distributed task execution service\SLP{db}, worker pools\SLP{workpools}, and a data sharing service\SLP{datasharing}.

% \begin{enumerate}
%     \item Overview diagram + lead in text

%     \item EMEWS DB: fault tolerance, Python and R API, stress the importance of our queue-based, multi-language focus
%     \item Async API: better utilization of HPC resources, avoid batch synchronous workflows, reprioritize, cancel; Code examples; currently developed in Python API and will be extended to R API
%     \item Worker pool: basic communication, batch thresholds etc., configuration/heterogeneous, scalability, MPI tasks
%     \item funcX for remote access, launch etc., indicate that there is the functionality
%     \item federated: connecting to special hardware resources, bursting capabilities, migrating tasks across resources as needed for robust workflows
% \end{enumerate}

\subsection{The Model Exploration Algorithm Module}
\label{sec:algmodule}

The central control of the OSPREY workflow is in the logic of the Model Exploration (ME) algorithm\circled{1}. This is the main user interface to the system.  The users deploy scientifically-oriented algorithms here, using a wide range of supported languages, including Python and R.  This allows users to draw from the extensive set of community-developed data science and analysis tools available. Such algorithms typical provide a pluggable interface in which to provide user functions to sample the scientific problem space. This is where the user connects the algorithm to the OSPREY task API.

The API for task submission, result reporting, and querying the queues is implemented in both Python and R, although the Python API has additional code that enables asynchronous ME algorithms\SLP{async}.  A task is submitted with the following arguments: an experiment id; the task work type; the task payload; an optional priority that defaults to 0; and an optional metadata tag string. The payload contains sufficient information for a worker pool to execute the task and is typically a JSON formatted string, either a JSON dictionary or in less complex cases a simple JSON list. On submission, the API creates a unique task identifier (an integer) for the task and inserts that identifier, the experiment identifier, the work type, and the payload into the EMEWS DB \SLP{db} tasks table, together with a task creation timestamp. That task identifier, priority and work type are then inserted into the EMEWS DB output queue table.

\subsection{Task Distribution}
\label{sec:tasks}

Tasks produced by the ME algorithm are distributed over the wide area network via a configurable network, with funcX~\cite{Chard2020funcX:} or SSH as the transport mechanism~\circled{2}.  Overall, the task distribution system decouples the tasks produced by the ME algorithm, and the status of those tasks (queued, running, etc.) from the ME execution such that tasks and their results are not lost when a resource fails, but rather are described in the system in enough detail so that they can be executed if not yet running or restarted if necessary.

We have built the OSPREY computational fabric upon the Argonne-developed funcX platform~\cite{Chard2020funcX:}. FuncX implements a federated function as a service (FaaS) model via which arbitrary Python functions can be reliably executed on remote computers. Users first deploy specialized funcX endpoint software on a computer to make it accessible for remote computation. The hosted funcX cloud service acts as an interface for users to submit tasks. The service is responsible for managing secure communication with an endpoint, authenticating and authorizing users (via OAuth 2.0), providing fire-and-forget execution by storing and retrying tasks in the event an endpoint is offline or fails, and storing results (or failures) until retrieved by a user. 
The funcX endpoint is responsible for provisioning resources via various supported systems (e.g., local fork, Slurm, PBS), managing execution of tasks using a pilot job model, optionally deploying tasks within containers, monitoring execution, and returning task results to users via the cloud service.

In our prototype, we use funcX to start and stop the EMEWS service, the EMEWS DB database \SLP{db}, and remote worker pools on HPC resources \SLP{workpools} \circled{3}.  The EMEWS service is a Python application and can thus be started directly from within a Python function executed on a remote funcX endpoint. The database and worker pools are non-Python applications, and those are started using funcX and the Python subprocess library to start the relevant executables. Due to the flexibility of  funcX, we can also use it to execute simple, standalone tasks on remote resources as need, such as for initiating large data transfers, or, as we show in \SL{results}, for an optimization phase in a Python ME algorithm that is best executed on a specific HPC resource.

\subsection{EMEWS Task Database}
\label{sec:db}

Tasks arrive at HPC sites at the EMEWS Service \circled{4}, which abstracts task caching and queuing operations in an efficient manner.  The Service stores the tasks in the EMEWS DB, a resource-local SQL database. The Service mediates between model exploration algorithms and worker pools and exposes data about tasks for queries.

% EMEWS DB is a multi-language package that, together with a relational database, mediates between model exploration algorithms and worker pools. MEs submit tasks (e.g., sets of simulation input parameters to be evaluated) to the database and worker pools query the database, retrieving those tasks for execution, and reporting the results of those tasks back to the database where the ME can then retrieve them.

The EMEWS DB consists of five tables, a tasks table, a table each for an input and output queue, an experiments table that links tasks with experiments, and a tag table that links individual tasks to a metadata tag or tags. The tasks table contains the data for each task: a unique task identifier; current status (queued, running, complete, or canceled); a task work type; the input payload (e.g., simulation input parameters); the result payload; the identifier of the worker pool running the task; the time the task was created; the task execution start time; and the task execution stop time. The output queue table, from which tasks are popped for execution, consists of a task identifier, a task work type, and a priority that determines the order in which the tasks should be executed. The input queue table consists of a task identifier and a task type. The rows in the tables are linked through the task identifier such that tasks in the input and output queue and in other tables link to those in the tasks table via the shared task identifier.  This schema provides the foundation for a fault tolerant and robust task queuing and execution workflow.

Client code can query for tasks in the output queue by passing an integer specifying the type of work, the number of tasks to retrieve, an optional string identifying the specific worker pool consuming the tasks, and an optional timeout and delay value. The query executes by polling the output queue table for tasks of the specified work type, using the specified delay and timeout value. On success, the highest priority task (or tasks if multiple tasks are requested) is deleted and returned from the output table, effectively popping the task off the output queue. The task's payload is retrieved from the task table which is updated, setting the task status to running, recording the start time, and setting the worker pool identifier to that passed in by the query. The payload and the task identifier are returned as a Python dictionary or a R named list: {\tt\string{\textquotesingle type\textquotesingle: \textquotesingle work\textquotesingle, \textquotesingle eq\_task\_id\textquotesingle: task\_id, \textquotesingle payload\textquotesingle: payload\string}}. If the polling fails, a dictionary or named list with a {\tt\textquotesingle type\textquotesingle} of {\tt\textquotesingle status\textquotesingle} and a {\tt\textquotesingle payload\textquotesingle} describing the reason for the failure (e.g., {\tt\textquotesingle TIMEOUT\textquotesingle}) is returned.

When the work defined by the {\tt\textquotesingle payload\textquotesingle} has been completed, the result, typically in JSON format, is reported and inserted into the input queue by passing the task identifier, the work type, and the result JSON payload. The result payload is inserted into the tasks table where the task status is marked as completed, and the stop time is recorded.  The task identifier and work type are inserted into the input queue table, effectively pushing the task into the input queue. Client code, such as an ME algorithm, queries for results by polling the input queue table for completed tasks, passing a task identifier together with timeout and delay values that work identically to those in the output queue query. On success, the task is popped off the input queue by deleting it from the input queue table, and the corresponding result from the tasks table is returned. On failure, an appropriate message is returned.

\subsection{Heterogeneous Worker Pools}  %% Or Extensible
\label{sec:workpools}

%Worker pool: basic communication, batch thresholds etc., configuration/heterogeneous, scalability, MPI tasks
In the previous section \SLP{tasks} we described how tasks can be submitted to, and how results are retrieved from the EMEWS DB.
Worker pools~\circled{5} are responsible for querying the database via the output queue for those submitted tasks,
executing them, and then reporting the results of those tasks back to the database where the ME algorithm can then retrieve 
them via the input queue. Our canonical worker pool implementation is a Swift/T~\cite{Wozniak2013Swift/T:} application 
written using the Swift language and Python, running on a HPC resource as a pilot job. The ability to manage an extreme quantity of tasks is a main design feature of the Swift/T implementation, which essentially distributes work among previously launched workers using MPI messages.  Swift/T is a dataflow language with built-in concurrency designed for execution on large-scale supercomputers. Our worker pool implementation leverages these
features to execute as many submitted tasks in parallel as possible. For performance results of epidemiologic models running on Swift/T workflows see ~\cite{Ozik2021population}. 

Swift/T also provides high-level, easy to use interfaces for Python, R, Julia and Tcl, allowing the user to pass a 
string of code into the language interpreter for execution. This feature enables us to integrate Swift's concurrent capability
with the Python EMEWS DB API for querying the output queue for tasks to execute, and for pushing the results of executed tasks on to the 
input queue. We have also implemented an enhanced version for querying the output queue,
customized for worker pools. These queries allow a worker pool to request up to $n$ number of tasks (a query \emph{batch size}) to consume at a time, while accounting for the number of tasks a worker pool already has obtained but have not completed. So, for example, if a worker pool is configured to possess 33 tasks at a time, if it owns 30 uncompleted tasks when querying the output queue, it will only obtain 3 additional tasks. This can be tweaked using a \emph{threshold} value that specifies how large the
deficit between requested tasks and owned tasks must be before more tasks are obtained. Querying for tasks in this way
allows a worker pool to tune its query to the number of available workers such that all its workers are busy while equitably
sharing work among multiple worker pools. This prevents any one worker pool from obtaining more tasks than it can reasonably execute while
potentially leaving other pools starved of work.

Each worker pool has a user specified work type associated with it when it is initialized. A pool will only query for and execute tasks
of that type. Consequently, worker pools can be matched to HPC resources configured to most efficiently execute their work type.
For example, an ME algorithm may have two types of tasks that need to be executed: 1) a multi-process MPI-based simulation model;
and 2) an optimization component that most efficiently runs on a GPU. Two worker pools can be launched and configured
on resources appropriate for these two different work types. Worker pools also need not be started and stopped
when the ME algorithm starts and stops. They can be started and stopped as needed in response to changing ME algorithm requirements. 

Worker pools can run a variety of task application types.  In addition to the Python, R, Julia, and Tcl interfaces which can be used to run
code written in those languages, Swift/T can run executables as command line programs via its {\tt app} function type. MPI
executables can also be run via an {\tt app} function or internally as parallel Swift/T library extensions using
the {\tt @par} keyword. 

\subsection{Data Sharing Service}
\label{sec:datasharing}

One of the challenges with our approach is efficiently
moving data to/from the HPC resources. Staging data (including models) via, e.g., SCP requires authentication (often two-factor authentication
on HPC resources), while using funcX is not possible
as funcX limits input/output sizes to 10MB. 
To address the need for out-of-band transfer of potentially
large data, we use ProxyStore~\cite{ProxyStore_2022} and Globus~\cite{Chard2014Efficient}.
This communication is shown in Figure~\ref{fig:overview}~\circled{6} and allows for seamless access to remotely-hosted large data sets across HPC sites, with no changes to code.

% Text from proposal that talks about ProxyStore and Globus
% Traditionally, the change in technologies used from wide-area data management to HPC-oriented data manipulation are a challenge for workflow development. We will develop performant and uniform access to data irrespective of where a task executes in the OSPREY ecosystem. Typical OSPREY workflows will require a usable interface to referencing, accessing, and storing data as well as a robust backend for moving data both among HPC resources and within HPC resources. We will build upon two systems developed by our teams: ProxyStore~\cite{ProxyStore_2022} and Globus~\cite{Chard2014Efficient} to support these needs.

ProxyStore is a data management fabric that exposes a simple (and common) Python interface to data irrespective of where it resides. 
Specifically, it passes ``Proxy'' object references between participating entities (e.g., ME algorithms, EMEWS DB, remote workers) and implements a lazy evaluation approach in which Proxies are resolved only when needed. Thus, users are presented with a pure Python interface which can be easily integrated with various models and client software. ProxyStore implements a common data access/movement interface with plugins to support storage and movement via different methods, including shared file systems, Redis databases, or Globus.  

Globus provides high-performance and reliable third-party data transfer and is available on HPC systems in national laboratories, national cyberinfrastructure, and in research institutions. It can also be easily deployed on laptops and clouds. The third-party nature of Globus transfers, allows OSPREY (via ProxyStore) to easily move data between locations without needing to maintain open connections to those locations. 

\vspace{3mm}

The capabilities of the OSPREY prototype architecture described, including the ME algorithm \SLP{algmodule}, the control and task distribution system\SLP{tasks}, the  distributed task execution service\SLP{db}, the worker pools\SLP{workpools}, and the data sharing service\SLP{datasharing}, provide the ability to robustly and securely coordinate workflows across a distributed ecosystem of heterogeneous remote resources \circled{7}.

% We will experiment with Globus as the primary backend for data movement, using a batch-based model to collect data into fewer Globus transfer operations. We will extend, develop, and integrate MPI-based techniques~\cite{Wozniak2014Big} with ProxyStore to rapidly move data into node-local storage for fast movement on a site and integration with advanced data features like DAOS~\cite{DAOS_WWW}, will enable extremely fast data sharing within an HPC site.

\section{Programming Model and APIs}
\label{sec:apis}

\subsection{Task model}

Examples of the base API for submitting tasks, querying for tasks to execute,
reporting results, and querying for results in both Python and R can been seen 
in Listing \ref{lst:base_api}.
The Python version currently has some additional functionality, the ability to query for
multiple tasks to execute, for example, which accounts for differences in the
method and function signatures. Further, the Python API is class based in which 
instances of an {\tt EQSQL} Python class provides methods for submission, querying and reporting.

    \begin{small}
    \begin{lstlisting}[caption=Core EMEWS DB API in Python and R., label=lst:base_api]
# Python API
def submit_task(self, exp_id: str,
    eq_type: int, payload: str, 
    priority: int = 0,
    tag: str = None)

def query_task(self, eq_type: int, n: int = 1,
    worker_pool: str = 'default',
    delay: float = 0.5,
    timeout: float = 2.0)

def report_task(self, eq_task_id: int,
    eq_type: int, result: str)

def query_result(self, eq_task_id: int,
    delay: float = 0.5,
    timeout: float = 2.0)

# R API
eq_submit_task <- function(exp_id, eq_type,
    payload, priority=0)

eq_query_task <- function(eq_type, delay = 0.5,
    timeout=2.0)

eq_report_task <- function(eq_task_id,
    eq_type, result)

eq_query_result <- function(eq_task_id,
    delay = 0.5, timeout = 2.0)\end{lstlisting}
    \end{small}

\subsection{Asynchronous Tasks}
\label{sec:async}

In section  \ref{sec:fast-time}, we discussed the importance of asynchronous algorithms for fast
time to solution, and for providing better utilization of HPC resources when compared with
batch synchronous workflows. Here we describe the OSPREY asynchronous API that
enables these algorithms. The asynchronous API is designed using the {\em future} abstraction.
A future encapsulates the asynchronous execution of a task, and is implemented as a Python
class. Future instances are created and returned when tasks are submitted with {\tt EQSQL.submit\_task}.
Leveraging the EMEWS DB API, {\tt Future} class methods allow ME algorithm code to query 
the status (running, finished, etc.) and check for a result of the encapsulated task without waiting for it to finish. Other methods provide the ability to cancel and reprioritize the task with 
respect to other tasks in the output queue. 

The asynchronous API also includes functions for working with collections of {\tt Futures}.
Given a list of {\tt Futures}, the {\tt as\_completed} function creates a Python generator
that will yield a specified number of {\tt Futures} as they complete. {\tt pop\_completed} 
returns the first completed {\tt Future} from a list of {\tt Futures}, removing that 
{\tt Future} from the list, and {\tt update\_priority}
will update the priorities of a list of {\tt Futures} to a new set of priorities. 
For efficiency, these functions typically perform batch operations on the EWEWS DB 
rather than iterating through the collection of {\tt Futures} and performing the operations individually.
Taken together, these functions and the {\tt Future} class methods enable the implementation of
various asynchronous algorithms.

The pseudo-code in Figure \ref{figure:async_psuedo_code} illustrates a typical asynchronous
algorithm. A number of initial samples are submitted for evaluation. After 
some number of evaluations have completed and their results are available, the ME algorithm, using these results,
can generate new samples for evaluation,
reorder existing evaluations, cancel less promising evaluations, and so on, until some stopping
condition is reached. The code in Listing~\ref{lst:async} is a possible implementation of the pseudo-code
using our asynchronous API. Lines 4 through 9 create JSON payloads for each sample in a list
of samples, submitting those for evaluation using the {\tt EQSQL} class instance created on line 2.
The {\tt Futures} returned by the submission are added to a {\tt futures} list in line 9. In line
13, the {\tt pop\_completed} function pops a completed future from the list. In this case, and by 
default, {\tt pop\_completed} will poll for a completed result until one is found. However,
a maximum wait time can also be specified using a {\tt timeout} argument. In line 15, an
{\tt update} function takes the result of the completed future, and generates new
tasks to be submitted, and new priorities for the existing tasks. Line 16
updates the existing tasks encapsulated by the {\tt futures} list to the new
priorities, and lines 17-20 submit the new tasks, producing new futures, which
are then added to the {\tt futures} list where they can be popped off as they complete.

\begin{figure}[tbp]
    \begin{algorithmic}[1]
    \For{each initial sample}
        \State Submit the sample for evaluation
    \EndFor
    \While{stopping condition not reached}
        \State Wait for $n$ number of evaluation results
        \State Re-sample, reorder, re-submit based on results
    \EndWhile    
\end{algorithmic}
\caption{Pseudo-code for an asynchronous algorithm.}
\label{figure:async_psuedo_code}
\end{figure}

% \begin{figure}[tbp]
    \begin{small}
    \begin{lstlisting}[numbers=left, caption=Sample asynchronous algorithm implementation., label=lst:async, upquote=true]
from eqsql import eq
eqsql_db = init_esql(...)
futures = []
for s in samples:
    payload = json.dumps({'sample': sample})
    ft = eqsql_db.submit_task(
        'exp1', sim_work_type, payload, 
        priority=0)
    futures.append(ft)

tasks_completed = 0
while tasks_completed < 1000:
    ft = eq.pop_completed(futures)
    tasks_completed += 1
    tasks, new_priority = update(ft.result())
    eq.update_priority(futures, new_priority)
    for t in tasks:
        ft = eqsql_db.submit_task(
            'exp1', sim_work_type, task)
        futures.append(ft) \end{lstlisting}
\end{small}
%\end{figure}

% Maybe too specific to Python here --- When a task requires something for its execution that cannot be insert into the database
% as a payload, we use the ProxyStore library {\bf REF} and its Globus {\bf REF} functionality. ProxyStore provides an interface to object stores through small lightweight proxy objects. Using ProxyStore, we can create a proxy object from some large object
% and use the proxy object as part of the task payload. ProxyStore will transparently copy the large object to the remote
% resource. The task 

% NOTE: This text is here in case it's useful. It currently is very funcX centric, which doesn't work with the section flow!
%While funcX provides a uniform interface for remote execution, it addresses a narrow section of the types of computations needed in OSPREY—single node Python functions. To address the needs of a wider range of tasks, we will build on funcX to enable it to launch tasks within extensible worker pools managed by other workload managers (e.g., Swift/T) for specialized work (e.g., multi-node MPI tasks). For these workloads, we will dynamically deploy funcX functions that themselves deploy worker pools using the mechanism prescribed by the workload manager. Subsequent execution of tasks will use the workload manager itself to execute tasks via the deployed worker pool.

\section{Results}
\label{sec:results}
To exercise our prototype OSPREY HPC architecture, we have implemented an example optimization workflow
that attempts to find the minimum of the Ackley function~\cite{ackley_func} using a 
Gaussian process regression model (GPR). Our implementation, which can be found at 
\url{https://github.com/NSF-RESUME/2023\_ParSocial\_OSPREY\_example},
is based on a similar example problem provided as part of the Colmena documentation~\cite{colmena_ackley}.
We begin with a sample set containing a number of randomly generated n-dimensional points. 
Each of these points is submitted as a task to the Ackley function for evaluation. When
a specified number of tasks have completed (i.e., that number of Ackley function evaluation results
are available), we train a GPR using the results, and 
reorder the evaluation of the remaining tasks, increasing the priority of those more
likely to find an optimal result according to the GPR. This repeats until all the evaluations complete.

% \begin{figure}[tbp]
%     \begin{algorithmic}[1]
%     \State Given a sample set $S$ of $n$ number of $d$ dimensional points,
%     \For{ each point $p$ in $S$}
%         \State Submit $p$ to the Ackley function for evalution
%     \EndFor
%     \State $N \gets 0$
%     \While{$N \neq $n}
%         \State $R \gets r$ number of evaluation results
%         \State $S \gets S - R$
%         \State Train a Gaussian Process Regression $GPR$ model using $R$
%         \State Reprioritize evaluation order of $S$ using $GPR$
%         \State $N \gets N + r$
%     \EndWhile    
% \end{algorithmic}
% \caption{Example workflow pseudo-code}
% \label{figure:repro_psuedo_code}
% \end{figure}

Our example workflow was executed locally on an M1 MacBook Pro in conjunction with the 
University of Chicago's Midway2 HPC cluster, the Laboratory Computing
Resource Center's Bebop HPC cluster at Argonne National Laboratory, and the Argonne Leadership 
Computing Facility (ALCF) Theta supercomputer. The EMEWS DB components and worker pools were run
on Bebop, and the GPR training was done on Midway2 or Theta depending on the run configuration.

The ME algorithm is a Python script running locally that begins by initializing a funcX client, and
then starting the
EMEWS DB, an initial worker pool, and the EMEWS service remotely on Bebop using funcX as
described in section~\ref{sec:tasks}. After initializing an SSH tunnel through which we communicate
with the EMEWS service, we create an initial sample set of 750 4-dimensional points, which
are submitted as tasks using the {\tt submit\_task} API function. The worker pool, running on Bebop,
pops these tasks off the database output queue, and executes the Ackley function using the point
data in the tasks' payload. (We have added a lognormally distributed ``sleep'' delay to the Ackley function implementation to increase the otherwise millisecond runtime and to add task runtime heterogeneity
for demonstration purposes.) On completion, the task results are pushed onto the database's 
input queue. While the worker pool on Bebop is executing the tasks, the local Python script,
waits for the next 50 tasks to complete at which time we perform the reprioritization. The completed 
tasks are 
popped off the list of futures returned by the submission using the {\tt as\_completed} API
function.

The reprioritization consists of retraining the GPR with the completed results and then updating the
evaluation priorities of the uncompleted tasks using the GPR predictions. The retraining of the GPR
was performed on Midway2 or Theta using funcX to directly evaluate the Python function that retrains
the GPR and returns the updated evaluation order. The GPR itself was passed as a ProxyStore proxy
object, using ProxyStore's Globus functionality, to the reprioritization function and resolved into the actual GPR during remote function evaluation.
Using the updated order returned from the function, the uncompleted tasks are reprioritized using the
{\tt update\_priorities} API function. The reprioritization repeats for every new 50 completed tasks, and start an additional worker pool
on Bebop after the 2\textsuperscript{nd} and 4\textsuperscript{th} 
reprioritizations, for a final total of 3 worker pools. Connecting to the same database as the initial worker pool, these worker pools perform
the same type of work, popping
tasks off the same output queue, and executing the Ackley function using those tasks' payload.
When there are no more tasks left to complete, the workflow terminates, stopping the database, and shutting
down funcX.

\begin{figure}[htbp]
    % \centerline{\includegraphics[width=0.5\textwidth]{figures/HPC_v6.png}}
  %% OSPREY-Arch-ME.pdf is at:
  %% https://docs.google.com/drawings/d/1BOfdyXkPckx85SulrOLY062lpTeDxA6Xv6AKu56q9Fc
  \centerline{\includegraphics[width=0.5\textwidth]{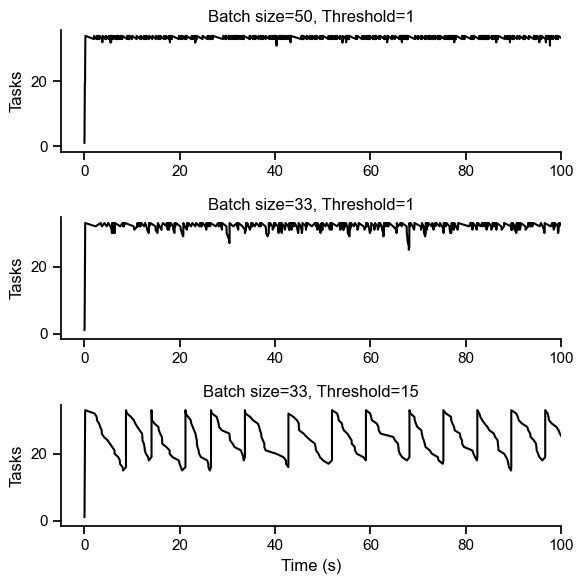}}
    \caption{Number of tasks executed by a worker pool for different batch sizes and thresholds.}
\label{fig:sawtooth}
\end{figure}

To illustrate the workflow's execution, we have created two figures. The first,  Figure~\ref{fig:sawtooth}, illustrates the effect of query batch size and threshold on worker pool utilization when querying for tasks. As mentioned in section \ref{sec:workpools} batch size controls the maximum number of tasks a worker pool can own and the threshold determines when additional tasks are requested. The top plot in the figure shows the number of concurrently running tasks with a query batch size of 50 and a threshold of 1 for a worker pool with 33 workers. The middle plot shows a batch size of 33 with a threshold of 1 for the same size worker pool, and the bottom plot a batch size of 33 with a threshold 15, again for the same size worker pool. The top plot clearly shows the best utilization of the HPC resource, in this case, a single 36 core compute node on Bebop. A batch size of 50 with 33 workers oversubscribes the pool, and effectively creates an in-memory task cache from which new tasks can be quickly pulled without the more costly database query. Oversubscribing, however, consumes database tasks, making them ineligible for reprioritization or cancellation, since they are popped off the output queue. With a batch size of 33 and threshold of 1 (middle plot), there is no such cache, and each time a task is completed another must be fetched from the database, during which additional tasks may complete, leading to lower utilization, but making more tasks eligible for reprioritization or cancellation. The final plot illustrates the effect of a large threshold where 15 tasks must finish before new tasks are added resulting in the saw tooth pattern where multiple workers remain idle for several seconds at a time.

\begin{figure}[htbp]
    % \centerline{\includegraphics[width=0.5\textwidth]{figures/HPC_v6.png}}
  %% OSPREY-Arch-ME.pdf is at:
  %% https://docs.google.com/drawings/d/1BOfdyXkPckx85SulrOLY062lpTeDxA6Xv6AKu56q9Fc
  \centerline{\includegraphics[width=0.5\textwidth]{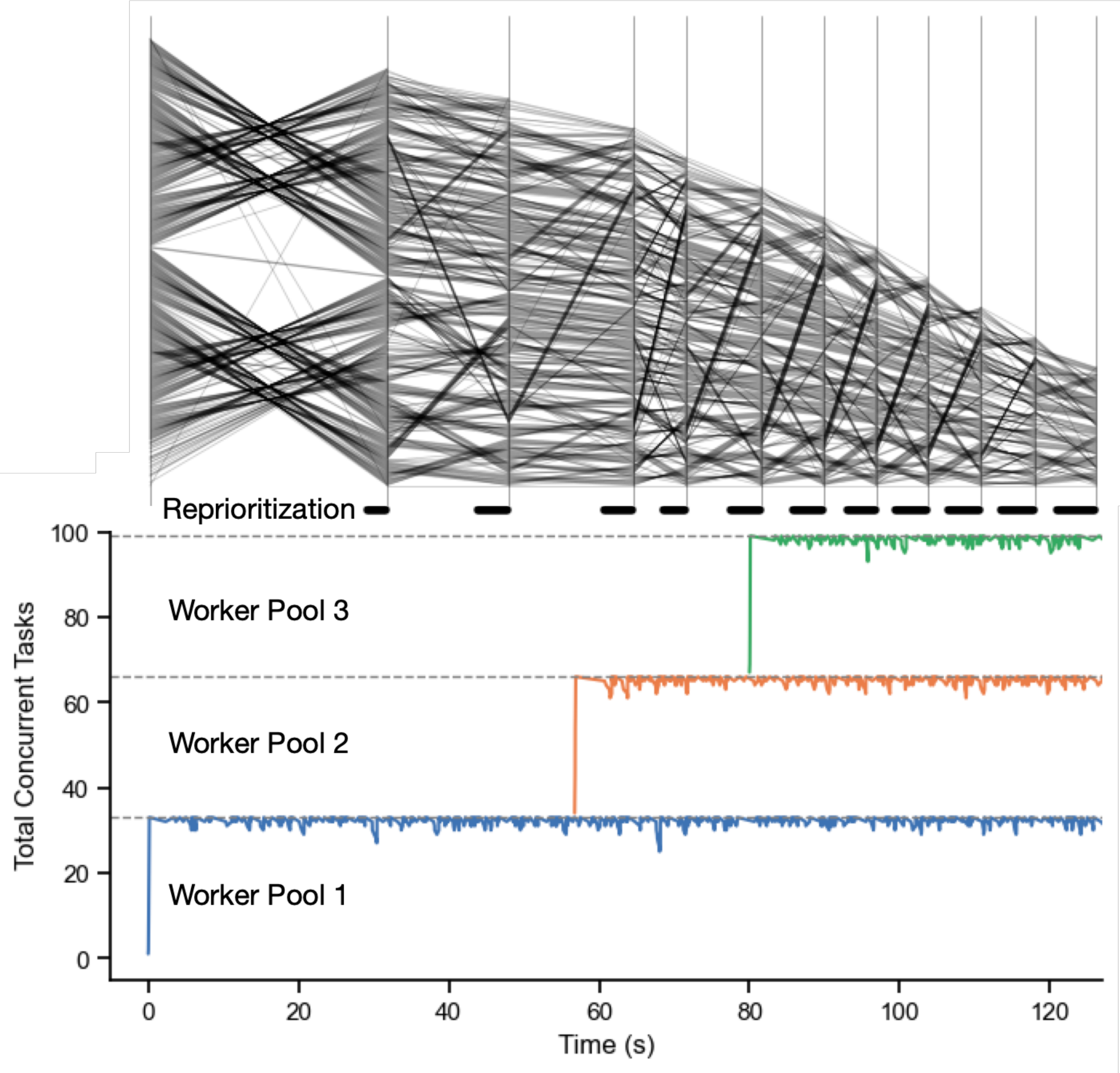}}
    \caption{Illustration of the combined example workflow across the ALCF Theta and LCRC Bebop resources. Top: GPR reprioritization of tasks over time, run on ALCF Theta. Vertical lines represent the new prioritization at the end of the reprioritization process, and connecting lines show task priority reassignments (higher means higher priority). As tasks are consumed, the tasks that are subject to reprioritization are reduced. The horizontal lines below the vertical lines show the time extent of each reprioritization call. Bottom: Total number of concurrently executing tasks by worker pool, run on LCRC Bebop.}
    \label{fig:tasks_reprio}
    \end{figure}

In Figure~\ref{fig:tasks_reprio} we have plotted the number of tasks executed by each of the three worker pools over time, together with the GPR reprioritization. The bottom of the figure shows the total number of concurrently executing tasks by worker pool, each of which has been allocated 33 workers and thus can execute up to that many tasks concurrently. In this example, each worker pool has a batch size of 33 and thus will query the database for up to 33 tasks at a time, and request more when the number of owned tasks falls below 33 (threshold of 1). In the figure, we see worker pool 1, in blue, starting at time 0 and begin executing 33 concurrent tasks, indicated by the first dotted line. When a task finishes, we see the total number of tasks drop below the maximum of 33 and then rise back up as more tasks are requested and executed. 57 seconds after worker pool 1 has started, worker pool 2, in orange, starts, and executes its maximum of 33 concurrent tasks. At the 80 second mark, worker pool 3 starts and begins to evaluate tasks. Across all the worker pools, we see a similar utilization to that in the middle plot in Figure~\ref{fig:sawtooth}. By using the batch API, we can equitably spread the tasks among the pools.

The top of the figure illustrates the dynamics of the GPR reprioritization occurring while
the worker pools consume tasks. The intermittent horizontal lines indicates the starting time and duration of the reprioritization with respect to worker pool task execution. We can see here how reprioritization becomes more frequent as the additional worker pools are added, given that 50 additional tasks complete more frequently with the additional worker pools. The very top of the figure shows the reprioritization trajectories of each of the 750 tasks, each line being drawn from a task's current priority to its new priority at the time of reprioritization. During the first reprioritization starting at the 29 second mark, and triggered by the first 50 tasks completing, 700 uncompleted tasks are reprioritized with new priorities of $1 - 700$, at the next reprioritization 650 uncompleted tasks are reprioritized from $1 - 650$, and so on. During reprioritization, the worker pools continue to consume tasks, efficiently using the compute resources, and providing the GPR with additional results to evaluate. Also of note, is that although worker pool 2 and worker pool 3 are scheduled to start during the 2\textsuperscript{nd} and 4\textsuperscript{th} reprioritizations, respectively they do not immediately start consuming tasks at that time due to delays between submitting a worker pool job to Bebop and it actually beginning.

% \begin{enumerate}
%     \item Intra worker pool dynamics
%     \item Reprioritizing algorithm setup
%     \item Swim lanes + reprioritizing graphic
% \end{enumerate}

\section{Conclusion and future work}
\label{sec:conclusion}

% COVID-19 had an unprecedented impact on scientific collaboration.  The pandemic and its broad response from the scientific community has forged new relationships among domain experts, mathematical modelers, and scientific computing specialists.  Computationally, however, it also revealed critical gaps in the ability of researchers to exploit advanced computing systems.  These challenging areas include gaining access to scalable computing systems, porting models and workflows to new systems, sharing data of varying sizes, and producing results that can be reproduced and validated by others.
% Informed by our team's work in supporting public health decision makers during the COVID-19 pandemic and by the identified capability gaps in applying high-performance computing (HPC) to the modeling of complex social systems, we present the goals, requirements, and initial implementation of OSPREY, an open science platform for robust epidemic analysis. The prototype implementation demonstrates an integrated, algorithm-driven HPC workflow architecture, coordinating tasks across federated HPC resources, with robust, secure and automated access to each of the resources. We demonstrate scalable and fault-tolerant task execution, an asynchronous API to support fast time-to-solution algorithms, an inclusive, multi-language approach, and efficient wide-area data management. The example OSPREY code is made available on a public repository.    

We have provided the goals and requirements for OSPREY, an open science platform for robust epidemic analysis. We described the prototype HPC OSPREY architecture and demonstrated an example of its use across federated HPC resources, providing the implementation code on a public repository. Our goal is to develop OSPREY into a public resource for epidemiologic modeling and analysis.

There are multiple directions for future work. To meet additional requirements in the integrated, algorithm-driven HPC workflow topic (see~\ref{sec:integ-hpc}), we will extend the asynchronous API to additional ME algorithm languages, starting with R, and expand the funcX capabilities for more robust interactions with HPC schedulers, including active monitoring and termination of worker pools, through the PSI/J library~\cite{Al-Saadi2021ExaWorks:}. For OSPREY capabilities in data ingestion, curation, and management (\ref{sec:req-data}), we will develop flexible techniques to provide real time and near-real time data to HPC workflows, automate the curation of the data, and provide methods for managing algorithm and model artifacts. Finally, to support the diffusion of OSPREY capabilities, we will develop the OSPREY shared development environment (\ref{sec:req-sde}), to promote model and workflow sharing, and to support best practices for model validation and reproducibility.
These capabilities will allow us to expand the applicability of OSPREY to diverse ME algorithms and integrated HPC workflows, with the purpose of creating on-demand response and planning capabilities to support public health decision making. 

Finally, while the focus of this work is in the public health domain, many aspects are relevant to other application areas where dynamic data and modeling on HPC resources can inform time-critical decision making.

\section*{Acknowledgment}
This research was completed with resources provided by the Research Computing Center at the University of Chicago (Midway2 cluster), the Laboratory Computing Resource Center at Argonne National Laboratory (Bebop cluster), and the Argonne Leadership Computing Facility (Theta), which is a DOE Office of Science User Facility.

%% NOTE: Used https://rintze.zelle.me/ref-extractor/ to extract refs from OSPREY doc.

\bibliographystyle{IEEEtran}
\bibliography{IEEEabrv,bib/osprey}

% \vspace{12pt}
% \color{red}
% IEEE conference templates contain guidance text for composing and formatting conference papers. Please ensure that all template text is removed from your conference paper prior to submission to the conference. Failure to remove the template text from your paper may result in your paper not being published.

\end{document}